\documentclass[11pt]{article}
\usepackage{moriond_cjl,epsfig}

\def\Journal#1#2#3#4{{#1} {\bf #2}, #3 (#4)}

\def\NIMA{{\em Nucl. Instrum. Methods} A}

\def\PLB{{\em Phys. Lett.}  B}

\def\be{\begin{equation}}
\def\ee{\end{equation}}
\def\bea{\begin{eqnarray}}
\def\eea{\end{eqnarray}}

\begin{document}
\vspace*{4cm}
\title{B PHYSICS AT LEP AND SLD}

\author{ C.S. Lin }

\address{Department of Physics, University of Massachusetts,\\
Amherst, MA 01003, USA}

\maketitle\abstracts{
Recent B physics results from LEP and SLD are reviewed.  
In particular, results of B lifetimes, semileptonic B decays,
charm counting, CKM matrix elements and mixing are presented.}

\section{Introduction}
LEP and SLD offer favorable environments to study the detailed properties
of the B hadrons.  
The clean $e^+ e^- \rightarrow Z^0$ environment along with
precision vertexing tools, allow one to easily identify B
hadrons, which have a typical decay length of about 3mm at the $Z^0$. 
The four LEP experiments have each accumulated 0.9 million $Z^0 \rightarrow
b\bar{b}$ events and the SLD experiment has roughly 120K $b\bar{b}$
events on tape.  Even with a small data size, SLD remains highly 
competitive in B physics due to the polarized $e^-$ beam, small beam spot 
and a unique pixel based CCD vertex detector.  These data sets have been
used in a broad B physics program and some recent results (lifetimes,
semileptonic decays, charm counting, $V_{ub}$, $V_{cb}$, and mixing) are 
presented in this paper.

\section{B Hadron Lifetimes}
The spectator model predicts that the lifetime
of a heavy hadron depends only on the properties of the decaying
heavy quark Q.  This model fails in the charm hadron sector for which
the lifetime hierarchy of $\tau_{D^+}>\tau_{D_s^+}>\tau_{D^0}>
\tau_{\Lambda_c^+}$
is observed.  Corrections to the spectator model can be calculated in the
context of the heavy quark expansion and are
predicted to scale with $1/m^2_Q$.  For the B hadrons, the
lifetime differences are expected to be less than 10$\%$.  Hence, precise 
measurements of B lifetimes provide a  test of our understanding 
of QCD in the heavy quark limit.  In addition, B hadron lifetimes are
important inputs to many heavy flavor analyses.

The most precise $B^+$ lifetime measurements are based on inclusive
topological techniques where charged secondary vertices are
selected to obtain a pure $B^+$ sample.\cite{workg}  
The $B_d^0$ lifetime has also been measured using the inclusive
method by selecting neutral vertices, although for the inclusive 
$B_d^0$ measurements, the background tends to be higher due to contamination
from the other neutral B hadrons ($B_s$ and B baryons).  
Recently, OPAL has made a measurement of $B_d^0$ lifetime using
$B_d^0 \rightarrow D^{*-} \it{l}^+ \nu$ decays.
The $D^{*-} \rightarrow D^0 \pi^+$ decays are reconstructed inclusively
by identifying the soft pions from the $D^{*-}$ decays.  This approach
yields the most precise $B_d^0$ lifetime measurement to date 
($\tau_{Bd}=1.541\pm0.028\pm0.023$ ps).\cite{taubd}  

\begin{figure}[h]
\center\psfig{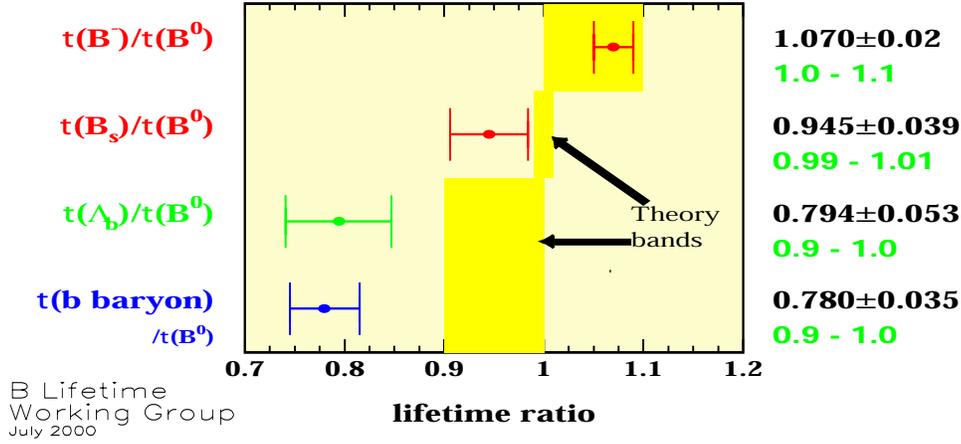}
\caption{Lifetime ratios vs. theory (yellow bands)\label{fig:tau_theory}}
\end{figure}

The lifetime measurements are most easily compared with theory through the 
lifetime ratios.  The measurements along with theoretical 
predictions~\cite{bigi} are shown in Figure~\ref{fig:tau_theory}.
The agreement is excellent in the meson sector.  However, the B baryon
lifetime measurements are consistently lower than predictions.  It remains
to be seen whether better theoretical predictions based on lattice
calculations can resolve the discrepancy.

\section{B Semileptonic Branching Ratio and Charm Counting ($n_c$)}
A long standing puzzle in B decay physics is the discrepancy between
the theoretical predictions and measurements of the B semileptonic branching 
ratio.  A simple solution to the puzzle is to increase the expected
rate of the non-semileptonic decay amplitude, thus lowering the
predicted semileptonic branching fraction.  There are also persistent 
disagreements among experiments on the value of 
Br($b\rightarrow X\it{l} \nu$).
More specifically, the measured values obtained at the $Z^0$ pole are
consistently higher than the ones obtained at $\Upsilon(4S)$.

The ALEPH collaboration has made a new measurement of the B
semileptonic branching ratio using two different techniques.
One is based on the lepton transverse momentum to distinquish 
prompt leptons from cascade charm leptons.  The other technique
uses the correlation between the charge of the lepton and the parent
quark charge.  The parent quark charge is determined by a charge
estimator built using tracks in the opposite hemisphere (e.g. jet charge).
Based on a sample of 886K candidates and an average B purity of about 90$\%$,
the measured semileptonic branching ratios are: 
Br($b\rightarrow X\it{l} \nu$)=0.1055$\pm$0.0009$_{stat}\pm$0.0024$_{syst}\pm
$0.0021$_{model}$ and 
Br($b\rightarrow{c}\rightarrow{X}\it{l}{\nu}$)=0.0804$\pm$0.0014$_{stat}\pm$
0.0024$_{syst}{\pm^{0.0009}_{0.0013}}$$_{model}$.\cite{absemi}

The SLD experiment has provided a new measurement of the charm production
rate from B decays using a topological vertexing method.
By fitting the number of reconstructed vertices per hemisphere and 
the vertex separation distributions 
to the Monte Carlo derived models, 
the following branching ratios are obtained: 
BR($b\rightarrow 0D$)=(5.6$\pm$1.1$_{stat}\pm$2.0$_{syst}$)\% and 
BR($b\rightarrow 2D$)=(24.6$\pm$1.4$_{stat}\pm$4.0$_{syst}$)\%.
Assuming that the
charmonia production rate is (2.4$\pm$0.3)$\%$ and that the remaining rate is
attributed to single charm hadron production, the average number of
charm quarks produced per B decay is calculated to be 
$n_c$=1.238$\pm$0.055.\cite{aschou}  
The new $n_c$ value is higher
than the previous world average and it is in the region preferred by
theory.

\begin{figure}[ht]
\vskip -0.4cm
\center\psfig{figure=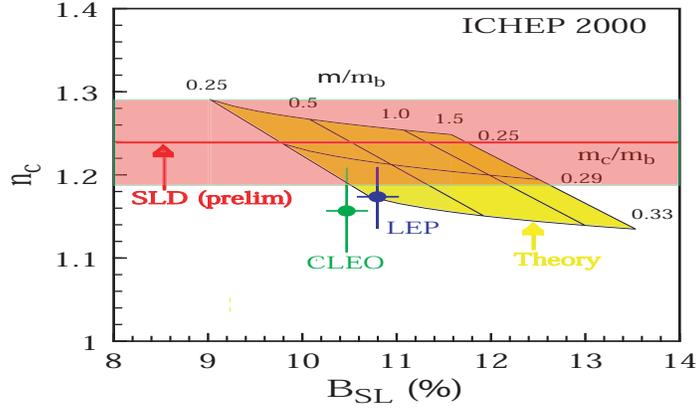,height=2.1in,width=3.6in}
\vskip -0.2cm
\caption{Semileptonic B.R. vs. $n_c$.  Yellow box is the theoretical 
predictions.
\label{fig:bslnc_sld}}
\vskip -0.2cm
\end{figure}

The current world situation is illustrated in Figure~\ref{fig:bslnc_sld}
with the theoretical predictions shown in yellow.
With the latest measurements and more recent calculations
which included higher order QCD perturbative corrections~\cite{neubert}, 
the experiment and theory are converging.  
The measured values from $Z^0$ and $\Upsilon(4S)$ 
are also now in good agreement.

\section{The Unitarity Triangle}
The major challenge in B physics for the coming decade is to perform
precision measurements of the CKM matrix elements and to test the Standard
Model description of CP violation.  The CKM matrix elements are required to 
satisfy the unitarity conditions such as  
$V_{ud}V^*_{ub}+V_{cd}V^*_{cb}+V_{td}V^*_{tb}=0$.  This relation is
commonly presented as a triangle in a complex plane with the apex of
the triangle denoted by the Wolfenstein parameters $\rho$ and $\eta$.  
The experimental goal is to over-constrain the triangle by measuring 
all three sides and three internal angles.  Any detected inconsistency in
the triangle relation may be a hint of new physics.   

\subsection{$|V_{cb}|$ and $|V_{ub}|$}
LEP uses both inclusive and exclusive 
approaches to measure the magnitude of $V_{cb}$.\cite{workg}  For the inclusive
method, the partial width for semileptonic B decay to charm mesons
is related to $|V_{cb}|$ using the expression:
$|V_{cb}|=0.044\times 
\sqrt{(Br(b\rightarrow c\it{l}\nu)\times1.55ps)/
(0.105\times \tau_b)}$.
In the exclusive method, the value of $|V_{cb}|$ is extracted by
measuring the differential decay rate for 
$B_d^0 \rightarrow D^{*-} \it{l}^+ \nu$ as a function of $\omega$, where
$\omega$ is the four-momentum product of $D^{*-}$ and $B_d^0$.  The
differential decay rate for the decay is given by:
$\frac{d \Gamma}{d \omega}=K(\omega)\it{F}^2(\omega)|V_{cb}|^2$, where
$K$ is the kinematic phase space term and $\it{F}$ is the hadronic 
form factor of the decay.  The unknown term in the expression is the form
factor, however, it can be calculated in the heavy quark limit
at $\omega$=1 which corresponds to the scenario where the $D^{*-}$ 
is at rest in the $B_d^0$ rest frame.  Experimentally, one measures the
differential decay rate near the kinematic endpoint and extrapolates to
$\omega$=1 to obtain the value of $|V_{cb}|$.  Results from the
inclusive and exclusive measurements are in good agreement.  The
combined results of the four LEP experiments gives 
$|V_{cb}|=0.0404\pm0.0018$. 

ALEPH, DELPHI, and L3 collaborations have recently extracted the value
of $|V_{ub}|$ from the measurements of 
$Br(b\rightarrow u\it{l}\nu)$.\cite{workg}
The expression relating $V_{ub}$ to the semileptonic $b\rightarrow u$
transition has been derived in
the context of the heavy quark expansion and is given as follows:
$|V_{ub}|=0.00445\times 
\sqrt{(Br(b\rightarrow u\it{l}\nu)\times1.55ps)/
(0.002\times \tau_b)}$.  The inclusive method has the
advantage of larger statistics and less model dependence compared 
to the exclusive and lepton endpoint analyses.  The combined average
of the three analyses gives $|V_{ub}|=0.0041\pm0.0007$. 

\subsection{B Mixing}
Measuring the $B_d^0$ oscillation frequency is the most direct way to
extract the CKM matrix element $|V_{td}|$.  The oscillation
frequency, $\Delta m_d$, has been measured to within
a few percent.  However, the extraction of $|V_{td}|$ from
$\Delta m_d$ is severely limited by theoretical uncertainties
of about 20\%.  The problem can be circumvented if one measures
the ratio of $\Delta m_d$ and $\Delta m_s$, where $\Delta m_s$ is the
oscillation frequency in the $B_s^0$ system.  As shown in
equation~\ref{equ:bsx}, in the ratio, many
theoretical uncertainties cancel and that allows one to extract 
$|V_{td}|$ to about 5\%.  
\begin{equation}
\label{equ:bsx}
{\Delta m_s\over\Delta m_d}={m_{B_s} f_{B_s}^2B_{B_s}\over 
m_{B_d} f_{B_d}^2B_{B_d}}|{V_{ts}\over V_{td}}|^2=
(1.15\pm0.05)|{V_{ts}\over V_{td}}|^2
\end{equation}

Experimentally, three basic ingredients are needed to study the
time dependent $B_s^0$ oscillations: 
1.) select a sample enriched in $B_s^0$ decays,
2.) determine the flavor of the B meson at production and decay and 
3.) recontruct the proper decay time.
The $B_s^0$ oscillation frequency is expected to be large in the Standard Model
and no direct observation (maybe a hint) has been established yet.
Nevertheless, a limit on $\Delta m_s$ is a strong constraint on the
apex of the Unitarity triangle (a lower limit on $\Delta m_s$ translates to
an upper limit on $|V_{td}|$).

\begin{figure}[h]
\vskip -.3cm
\center\psfig{figure=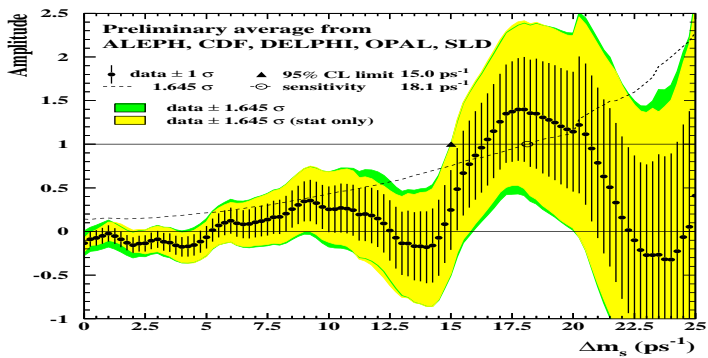,height=2.0in,width=4.0in}
\vskip -0.4cm
\caption{World amplitude plot showing the combined $B_s^0$ oscillation
amplitudes as a function of $\Delta m_s$.\label{fig:ampfit}}
\vskip -0.5cm
\end{figure}

LEP and SLD have carried out complementary programs to search for
$B_s^0$ oscillations.  A wide variety of techniques from the
fully inclusive SLD charge dipole to the fully exclusive $B_s^0$
reconstruction at ALEPH and DELPHI have been studied.  The
results have been combined based on the amplitude fit method.\cite{moser}
The world amplitude plot combining LEP, SLD and CDF results is shown
in Figure~\ref{fig:ampfit}.  The result excludes $\Delta m_s <$15 $ps^{-1}$
at the 95\% confidence level.\cite{bosc} The most significant deviation from
the amplitude of zero
occurs at about 17 $ps^{-1}$.  The significance of the structure has been
evaluated using toy Monte Carlo and is estimated to be consistent
with a fluctuation at the 3\% level.

\section{Conclusion}
LEP and SLD have made significant contributions in B physics over
the past few years.  The topics discussed in this paper 
represent only a small fraction of heavy flavor physics that have
been studied at the $Z^0$ pole.  As this paper is being written,
many results are still being finalized.  The B physics era
at LEP and SLD is not over yet.  Stay tuned.


\section*{References}
\begin{thebibliography}{99}

\bibitem{workg}ALEPH, DELPHI, L3, OPAL, CDF, and SLD Collaborations, 
 CERN-EP-2000, SLAC-PUB-8492, hep-ex/0009052.

\bibitem{taubd}G.~Abbiendi {\it et al.} [OPAL collaboration],
  \Journal{\PLB}{493}{266}{2000}, hep-ex/0010013.

\bibitem{bigi} G.~Bellini, I.I.~Bigi and P.J.~Dornan, 
  Phys. Rep. {\bf 289} (1997).

\bibitem{absemi}N.~Marinelli {\it et al.} [ALEPH collaboration],
  ALEPH 2000-069 (2000).

\bibitem{aschou}K.~Abe {\it et al.} [SLD collaboration], 
  SLAC-PUB-8686 (2001).

\bibitem{neubert}E.~Bagan, P.~Ball, B.~Fiol and P.~Gosdzinsky,
  Nucl. Phys. Lett B {\bf 351} 546 (1995); \\
  M.~Neubert, C.T.~Sachrajda, Nucl. Phys. B {\bf 483} 339 (1997).

\bibitem{moser} H.G.~Moser and A. Roussarie,
 \Journal{\NIMA}{384}{491}{1997}.

\bibitem{bosc}B Oscillations Working Group 2000, 
 http://www.cern.ch/LEPBOSC/

\end{thebibliography}
\end{document}